%% file: main.tex
\documentclass[journal]{IEEEtran}

\usepackage{graphicx}
\graphicspath{ {./figures/} }
\usepackage{subfig}
\usepackage[font=footnotesize]{caption}
\usepackage[table, x11names]{xcolor}
\usepackage{makecell}

\usepackage{threeparttable}

\makeatletter
\newcommand*\titleheader[1]{\gdef\@titleheader{#1}}
\AtBeginDocument{%                                                              
  \let\st@red@title\@title
  \def\@title{%                                                                 
    \bgroup\normalfont\normalsize\centering\@titleheader\par\egroup
    \vskip1ex\st@red@title}
}
\makeatother

\title{Approximate Decision Trees For Machine Learning Classification on Tiny Printed Circuits}
\titleheader{\vspace{-12pt}To appear at the 23rd International Symposium on Quality Electronic Design (ISQED'22), April 6-7, 2022, Virtual.}

\begin{document}
\bstctlcite{IEEEexample:BSTcontrol}
% \title{Approximate Decision Trees For Machine Learning Classification on Tiny Printed Circuits}

\author{\IEEEauthorblockN{
Konstantinos~Balaskas\IEEEauthorrefmark{1}\IEEEauthorrefmark{2},
Georgios~Zervakis\IEEEauthorrefmark{2},
Kostas~Siozios\IEEEauthorrefmark{1},
Mehdi~B.~Tahoori\IEEEauthorrefmark{2}
and~J{\"o}rg~Henkel\IEEEauthorrefmark{2}
}\\
\IEEEauthorblockA{
\IEEEauthorrefmark{1}Aristotle University of Thessaloniki, Greece,
\IEEEauthorrefmark{2}Karslruhe Institute of Technology, Germany
}\\
\IEEEauthorblockA{
\IEEEauthorrefmark{1}\{kompalas, ksiop\}@auth.gr,
\IEEEauthorrefmark{2}\{konstantinos.balaskas, georgios.zervakis, mehdi.tahoori, henkel\}@kit.edu
\vspace{-10pt}}
}

\maketitle
\input{abstract}


\input{introduction}
\input{background}
\input{methodology}
\input{evaluation}
\input{conclusion}

\section*{Acknowledgment}
This work is partially supported by the German Research Foundation (DFG) through the project ``ACCROSS: Approximate Computing aCROss the System Stack'' HE 2343/16-1.

\nocite{*}
\bibliographystyle{IEEEtran}
\bibliography{ref.bib}

\end{document}

%% file: abstract.tex
\begin{abstract}
Although Printed Electronics (PE) cannot compete with silicon-based systems in conventional evaluation metrics, e.g., integration density, area and performance, PE offers attractive properties such as on-demand ultra-low-cost fabrication, flexibility and non-toxicity.
As a result, it targets application domains that are untouchable by lithography-based silicon electronics and thus have not yet seen much proliferation of computing.
However, despite the attractive characteristics of PE, the large feature sizes in PE prohibit the realization of complex printed circuits, such as Machine Learning (ML) classifiers.
In this work, we exploit the hardware-friendly nature of Decision Trees for machine learning classification and leverage  the hardware-efficiency of the approximate design in order to generate approximate ML classifiers that are suitable for tiny, ultra-resource constrained, and battery-powered printed applications.
\end{abstract}

%% file: introduction.tex
\section{Introduction}\label{sec:introduction}
Several application domains, including smart packaging or fast-moving consumer goods (FMCG) have experienced very limited infiltration by silicon-based computing systems.
The sub-cent cost requirements of such domains cannot be met by silicon technologies, which feature high manufacturing and testing costs.
Printed electronics (PE) based on maskless and additive processes have emerged to overcome this obstacle, which offer significant cost reductions~\cite{Mubarik:MICRO:2020:printedclf}.
Printed electronics additionally offer stretchability, flexibility and porosity benefits, which are infeasible for silicon devices~\cite{Bleier:ISCA:2020:printedmp}.
Note however, that PE elements are manufactured at the micrometer ($um$) scale and therefore cannot compete with their silicon-based counterparts (manufactured at nanometer scale) in terms of performance or power and area efficiency.
Though, low-voltage printed electronics are envisioned to enable battery-powered, or even self-powered devices for the aforementioned application domains~\cite{Mubarik:MICRO:2020:printedclf}.

Many of these application domains can be greatly enhanced by machine learning (ML) classification algorithms, such as Support Vector Machines (SVM), Decision Trees (DT), Random Forests (RF) etc.
Due to the low non-recurring engineering (NRE) costs and per unit-area fabrication costs of PE, the implementation of highly customizable on-demand printed bespoke ML classifiers, tailored for specific architecture and datasets on a given application, is feasible~\cite{Mubarik:MICRO:2020:printedclf}.
Bespoke ML architectures~\cite{Cherupalli:ISCA:2017:bespoke} offer considerable area reduction, since the trained parameters of the implemented model are hard-wired, thus simplifying the respective circuitry.
Note that constructing custom bespoke classifiers would be infeasible in silicon-based technologies.
Nevertheless, the implementation of ML classification architectures with PE elements faces several challenges, originating from the large feature sizes associated with PE (i.e., prohibitively large power and area overheads).
Hence, the implementation of ultra-low area bespoke ML classifiers using traditional design techniques is a taxing procedure and as a result, the study of bespoke ML classification algorithms for printed electronics is limited.

A promising solution for overcoming the aforementioned challenges is Approximate Computing (AC).
Approximate computing has been established as a design paradigm for achieving significant gains in several design metrics (e.g., area), by performing inexact computations for a given error-tolerant application~\cite{Han:ETS:2013}.
Existing works on approximate computing implemented on silicon technologies, generate hardware-efficient approximate designs by intelligently inserting approximations in the operation of exact circuits~\cite{Shafique:DAC:2015:gear, Lee:DATE:2017:prec1, Kim:TCASI:2020:prec2}.
Due to the increased complexity introduced by approximate design, recent efforts have proposed automated approximation frameworks~\cite{Zervakis:IEEEACC:2020, Balaskas:TCAS1:2021:aging}.

In this work, we utilize the principles of approximate computing to explore the area efficiency of ML classification algorithms for printed technologies.
More specifically, we focus on Decision Trees because of their proven ability to implement low-power and hardware-efficient bespoke circuits~\cite{Mubarik:MICRO:2020:printedclf}.
By leveraging AC techniques, we are able to trade-off negligible classification accuracy loss of bespoke ML architectures for large reduction in area overhead.
Thus, we intelligently explore the large design space of candidate pareto-optimal approximate solutions.
Our framework applies a dual approximation approach on the trained coefficients of Decision Trees, which represent threshold values for the inferred decision rules (i.e., comparator threshold values).
We jointly scale the precision of comparator input features and coefficients, and alter the threshold values towards area-friendly values.
This follows our observation that Decision Tree architectures can be implemented in a hardware-friendly manner by slightly modifying the precision and threshold of its comprising comparators. 
Substituting a threshold value with a hardware-friendly one in its vicinity achieves our goal of reducing the area of the bespoke classifier without sacrificing considerable classification accuracy.

To efficiently examine the large design space of possible bespoke Decision Trees, we adopt the well-documented NSGA-II~\cite{Deb:TEC:2002:nsga}.
Multiple approximate Decision Tree architectures are created throughout each generation of the genetic algorithm, and evaluated for their classification accuracy and area requirements, in order to discover optimal hardware-friendly solutions.
The generated approximate architectures push the boundaries of area and power efficiency beyond state-of-the-art bespoke ML implementations and are tailored for ultra-resource constrained PE systems.

We evaluate our framework over several bespoke DT architectures for multiple machine learning datasets~\cite{Dua:2019:datasets}.
Our experiments demonstrate that the generated pareto optimal approximate solutions deliver $3.2\times$ and $3.4\times$ area and power gains, respectively, for the conservative accuracy loss threshold of $1\%$.

The rest of the paper is organized as follows. Section~\ref{sec:background} provides background on printed electronics and bespoke ML architectures. 
Section~\ref{sec:methodology} presents our proposed framework for the generation of hardware-efficient approximate bespoke DT for tiny printed circuits. 
In Section~\ref{sec:evaluation} the experimental evaluation of our framework using state of the art EDA tools is described. 
Section~\ref{sec:conclusion} contains concluding remarks.

%% file: background.tex
\section{Background}\label{sec:background}

\subsection{Printed Electronics}\label{sec:pe}
Printed electronics (PE) refers to a fabrication technology which is based on printing processes, such as jet-printing, screen- or gravure-printing~\cite{cui2016printed}.
Due to the simple manufacturing process as well as low equipment costs, ultra low-cost electronic circuits can be fabricated, at drastically lower cost compared to silicon-based processes, which require expensive foundries and clean rooms, even with older technology nodes.

Rather than replacing silicon-based electronics, PE serves as a complement because it can not compete with silicon-based electronics in terms of performance, integration density and area. 
Hence, PE has several advantages:
it is possible to print on several rigid or flexible substrates, such as plastic foil or paper, and enables low cost production~\cite{Clemens2004, Perelaer2010}. 

\begin{figure}[ht]
    \centering
    \includegraphics[width=0.6\linewidth]{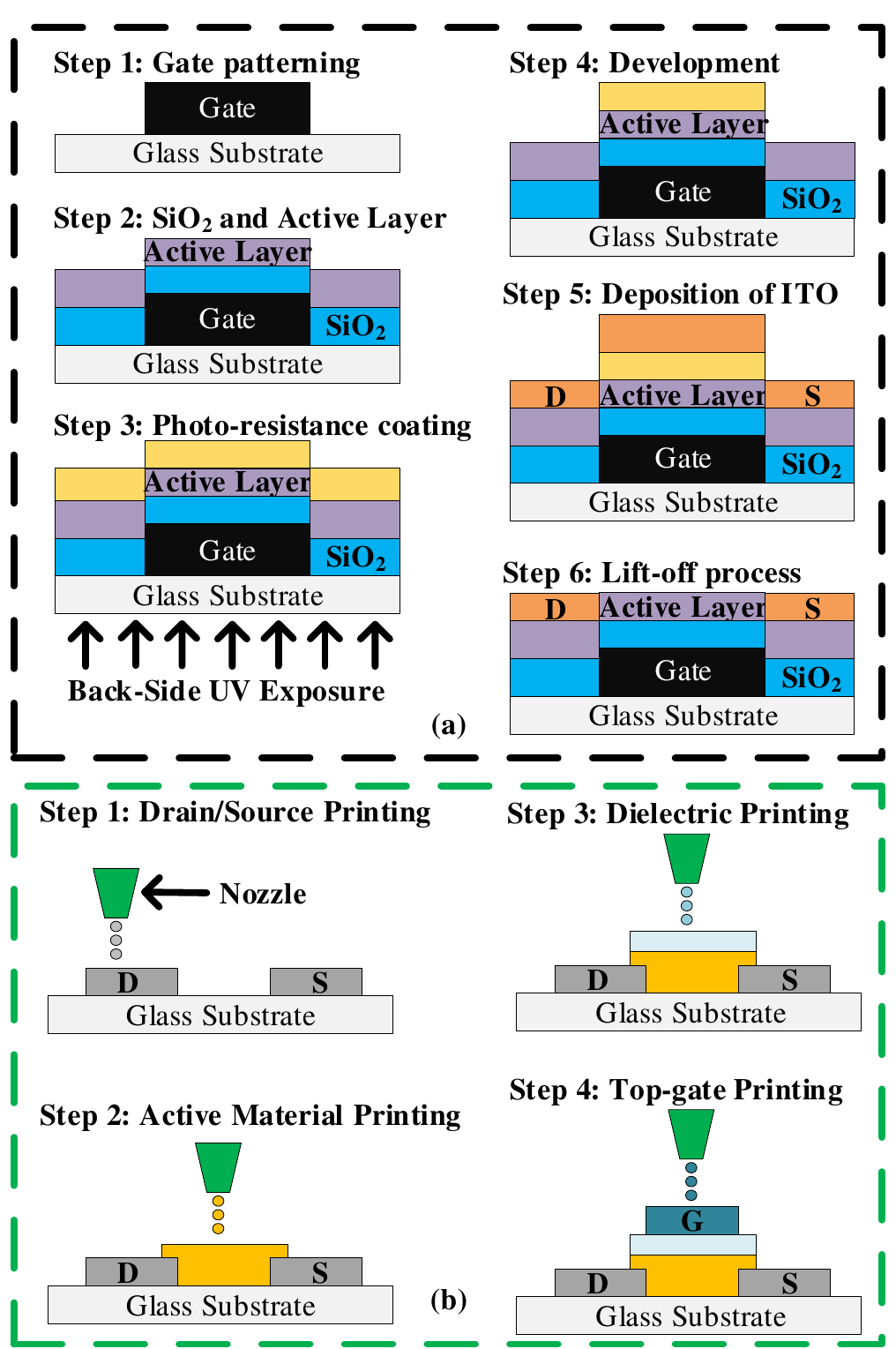}
    \caption{(a) Subtractive based printing process (b) Fully-additive inkjet printed process}
    \vspace{-10pt}
    \label{fig:printing_ovw}
\end{figure}

Printing technologies are broadly divided into two categories.
Some printing technologies are based on purely additive manufacturing process, while others employ subtractive process as well~\cite{chang2017circuits}, as shown in Fig.~\ref{fig:printing_ovw}.
In the subtractive process, a series of additive (deposition) and subtractive (etching) steps are involved, similar to silicon-based processing. 
The subtractive process is relatively expensive compared to the additive process, as it involves highly specialized processing, expensive equipment, and infrastructure. 
On the other hand, only deposition steps are involved in the additive manufacturing process.
Transistors, passive components, and interconnects are realized by depositing material layer-by-layer.
Generally, fully-additive printed electronics are slower compared to the subtractive-based printed technologies.

Electronics on flexible substrates are enabled by using contact-less printing methods, such as inkjet-printers, in combination with highly optimized functional inks such as conductive, semi-conductive and non-conductive materials. 
From these inks, organic~\cite{chung2011all} or oxide-based~\cite{shao2019recent} transistors can be built.
While organic materials are easy to be processed, they have lower environmental stability. 
On the other side, oxide-based inks have excellent conductivity and environmental stability, but are harder to be printed and suffer from impurities due to surfactants \cite{cui2016printed}. 

Inkjet-printed electrolyte-gated transistor (EGT) technology is an oxide-based inorganic printed technology, which deploys fluid electrolytes as a dielectric substitute in the transistor and allows operating voltages in the sub-$1V$ regime, making it a suitable candidate for self-powered portable computing systems in the IoT domain. 

Despite these promising features, still several limitations are prevalent in PE, which are large feature sizes and high parasitic capacitances, which lead to low functional densities and high device latencies. 
Due to this, low-complex circuit designs are favored with limited transistor count to reduce area utilization and to make the designs manufacturable with reasonable yield. 
Following this trend, several fundamental components for computing systems have already been successfully realized, such as boolean logic \cite{conti2020low}, digital and analog storage elements \cite{weller2018inkjet,huber2017fully}, or amplifiers \cite{kondo2018design}. 

\subsection{Bespoke ML Architectures}\label{sec:bespoke}
As aforementioned, printed electronics offer the possibility of implementing highly customizable ML classifiers, tailored for specific dataset and architecture, i.e., bespoke classifiers.
The elevated fabrication and NRE costs of silicon-based technologies prohibit the implementation of on-demand hardware with such high level of customization.
In bespoke ML classifiers, coefficients are hardwired in the circuit, leading to higher area efficiency than conventional circuits, compensating thus for the large feature sizes of PE elements and making the implementation of ML models in printed electronics feasible.
For example, the area of an unsigned 8-bit comparator (which is the basic building block for many ML classification algorithms) is on average $5$x larger than its bespoke implementations.
Furthermore, the logic of hardware components of bespoke architectures can be simplified by synthesis tools, since previously considered input parameters are set to constant hardwired values, which leads to further optimization and hardware efficiency.
The benefits of bespoke ML architectures for printed technologies were first quantified in~\cite{Mubarik:MICRO:2020:printedclf}, where both Decision Trees and SVMs were studied.
The authors implemented and evaluated bespoke printed classifer architectures, demonstrating their area efficiency and thus suitability for printed circuits. Moreover, \cite{Mubarik:MICRO:2020:printedclf} showed that, compared to other ML models, decision trees form perfect candidates for printed applications due to their lower area and power demands.

Our work focuses on bespoke Decision Trees (DT) under the scope of approximate computing and attempts to further improve on their hardware efficiency.
At the core of the DT structure, comparators learn decision rules to maximize the information gain from their input features.
We aim to implement approximate bespoke classifiers for ultra-resource constrained printed circuits, by approximating each comparator comprising a DT architecture, in an area-driven manner.
Compared to~\cite{Mubarik:MICRO:2020:printedclf}, we investigate ultra-low precision mixed-precision bespoke architectures (i.e. different precision for each comparator) to investigate the area-efficiency of approximate classifiers at a finer granularity.

%% file: methodology.tex
\begin{figure}[t]
    \centering
    \resizebox{1\columnwidth}{!}{
        \includegraphics{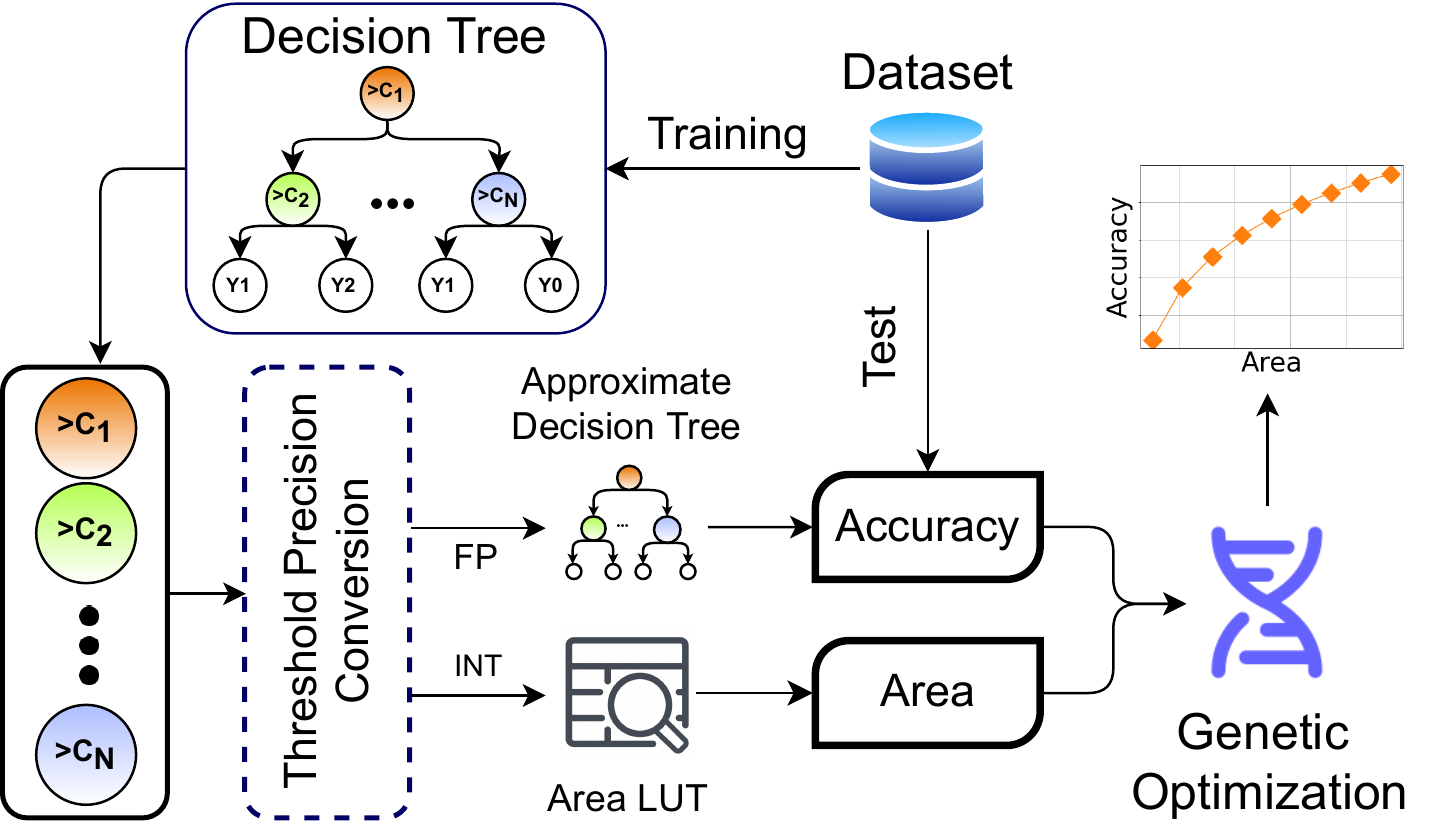}
    }
    \caption{Abstract overview of the proposed framework.}
    \vspace{-10pt}
    \label{fig:framework}
\end{figure}
\section{Design of Approximate Bespoke Decision Trees}\label{sec:methodology}
\begin{figure}
    \centering
    \subfloat[]{
    \resizebox{.7\columnwidth}{!}{\includegraphics{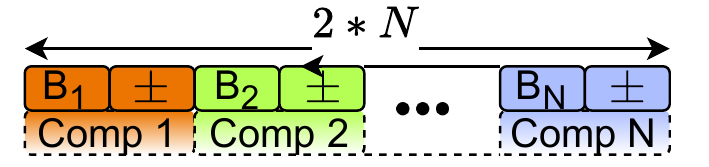}}
    \label{fig:chromosome}
    }
    \\
    \subfloat[]{
    \resizebox{.7\columnwidth}{!}{\includegraphics{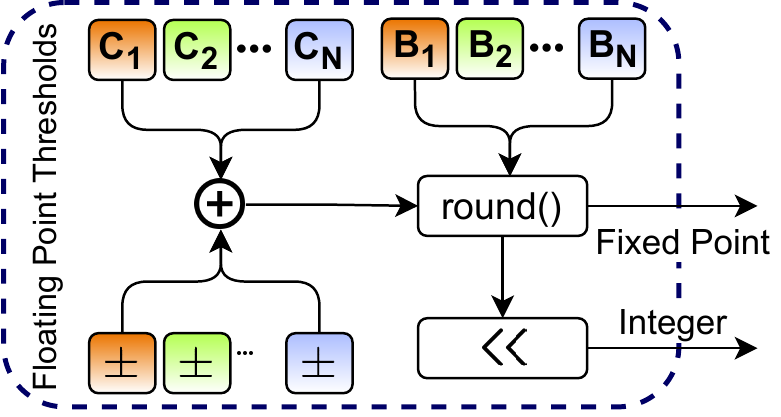}}
    \label{fig:conversion}
    }
    \vspace{-10pt}
    \caption{(a) Data format of chromosomes for each approximate bespoke decision tree. (b) Threshold precision conversion module (dotted blue box in Fig.~\ref{fig:framework})}
    \vspace{-10pt}
    \label{fig:chromosome_and_conversion}
\end{figure}

In this section, we present our automated framework for generating approximate hardware-efficient bespoke Decision Tree architectures for ultra-resource constrained printed devices.
Fig.~\ref{fig:framework} presents the abstract overview of our framework.
As input, our framework receives a trained Decision Tree model with a classification dataset.
Note that the dataset-agnostic nature of our implementation allows for arbitrary datasets and Decision Tree architectures to be examined.
From the comparators comprising the trained ML classifier, we obtain their threshold values in floating point precision.
We apply a dual approximation technique, by both scaling the precision of the input feature and threshold value of each comparator (i.e., mixed-precision quantization) and simultaneously applying an area-driven replacement technique on each threshold (see Section~\ref{sec:approximations}).
Approximations are facilitated by the conversion of floating point coefficients to fixed-point and integer, for accuracy and area estimation respectively, through our flexible theshold conversion module (Fig.~\ref{fig:conversion}).
To explore the vast design space of possible approximate solutions, we employ a genetic algorithm, like the one described in~\cite{Deb:TEC:2002:nsga}.

Our heuristic optimization generates close to pareto-optimal approximate DT designs in terms of area and classification accuracy.
The bespoke RTL description of the approximate solutions is automatically generated and synthesized using standard EDA tools and Process Design Kits (PDKs), specifically designed for printed technologies.
The generated gate-level netlists feature ultra-low area and power overheads, suitable for ultra-resource constrained and battery-powered printed applications.

\begin{figure}
    \centering
    \resizebox{1\columnwidth}{!}{
        \includegraphics{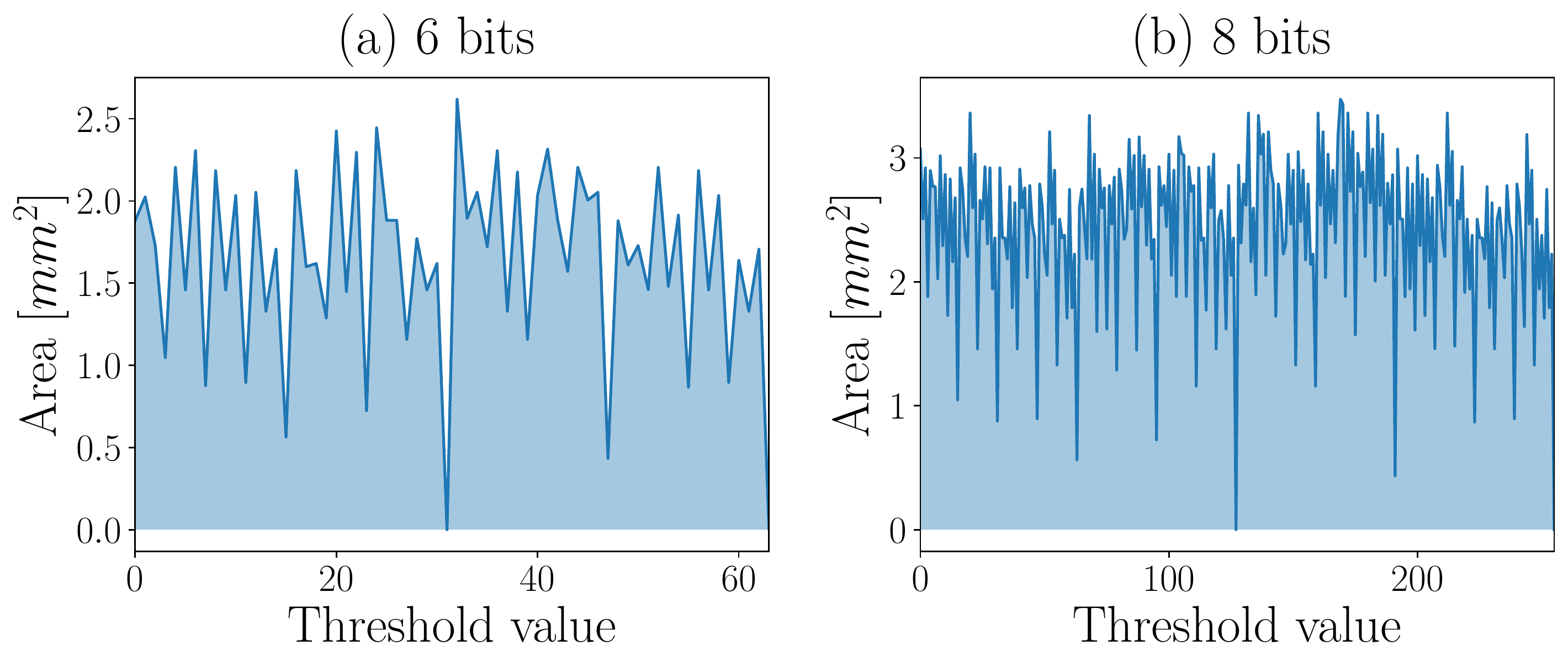}
    }
    \vspace{-10pt}
    \caption{Area measurements for bespoke comparators w.r.t. the threshold value. Two precisions are considered for inputs and thresholds: (a) 6 and (b) 8 bits.}
    \vspace{-10pt}
    \label{fig:area}
\end{figure}

\subsection{Comparator-Level Approximations}\label{sec:approximations}
In order to diminish the area and power demands of exact bespoke architectures, we employ a dual area-driven approximation methodology at the comparator level of Decision Trees.
Primarily, we observe that the area of bespoke comparators varies with its threshold value.
As previously mentioned, bespoke ML classifiers are designed with hardwired coefficients (i.e., threshold values in the case of Decision Trees) to enable high degree of customization and therefore hardware efficiency~\cite{Cherupalli:ISCA:2017:bespoke}.
Fig.~\ref{fig:area} presents the area measurements obtained from an exhaustive analysis of different integer threshold values of a bespoke comparator, with fixed precision at (a) 6 and (b) 8 bits for input features and coefficients.
It is evident that there is a non-linear dependency between the area and the threshold value of the comparator.
Thus, some coefficients are more hardware-friendly and result in an area-efficient implementation of bespoke comparators.

We leverage this observation to approximate each threshold value with a hardware-friendlier coefficient in its vacinity (i.e., within a predefined margin $\pm m$), leading to area gains for every comparator in the classifier.
Moreover, we scale the precision of the input feature and threshold value of each comparator.
Precision scaling is a well-studied approximation technique that removes the redundancy of floating point representations and can significantly reduce area and power overheads~\cite{Shafique:DAC:2015:gear, Lee:DATE:2017:prec1, Kim:TCASI:2020:prec2}.
Note, our framework applies precision scaling to every comparator in the Decision Tree, allowing for finer granularity than uniform precision and thus, further area reduction.

Our precision conversion module (Fig.~\ref{fig:conversion}) helps circumvent the mismatch of required representations for applying the two approximation techniques.
Precision scaling dictates the bitwidth of the fixed point threshold values, which are then converted to integers and replaced with hardware-friendlier substitutes, to comply with the integer-based findings of Fig.~\ref{fig:area}.
The fixed-point representation of the substituted and scaled coefficients is used to measure the classification accuracy on the selected dataset.
Applying both approximation techniques synergistically allows for significant area reduction over the exact bespoke DT architecture and enables the implementation of approximate classifiers on tiny printed circuits.

\subsection{Genetic Optimization}\label{sec:genetic}
Our automated framework employs an elitist, multiobjective, non-dominated sorting genetic algorithm (NSGA-II~\cite{Deb:TEC:2002:nsga}) to explore the vast design space of possible approximate bespoke Decision Trees.
The optimization procedure is guided by high level tools for estimating the accuracy and area of any given approximate solution.
Thus, we can fully exploit the inherently parallel nature of genetic algorithms to derive pareto-optimal approximate bespoke architectures in a fast and computationally efficient manner.
Measuring classification accuracy is straightforward (i.e., making class predictions on the test set), whereas area estimations would normally require synthesis of the RTL description of each approximate classifier.
Conducting synthesis on every approximate solution is time-consuming and restricts the parallel execution of our heuristic optimization to the number of available licences of the synthesis EDA tool.
To avoid this caveat, we utilize prior knowledge of area measurements from comparators and estimate the area of the Decision Tree as the sum of the area measurements of its comprising elements.
To that end, we store the comparator area measurements from our aforementioned exhaustive experiment (see Fig.~\ref{fig:area}) to create a look-up table (LUT) of area measurements for different input precisions  and integer coefficient values.
Thus, by correctly converting the threshold precision to integers (as seen in Fig.~\ref{fig:conversion}), area estimations of candidate approximate bespoke Decision Trees are seemlesly conducted at high level.

The approximation candidates for our two approximation techniques (see Section~\ref{sec:approximations}) are encapsulated in a chromosome, representative of each approximate decision tree, as seen in Fig.~\ref{fig:chromosome}.
It contains $2N$ genes, where $N$ is the number of comparators in the targeted bespoke classifier.
For every comparator, two genes (i.e., candidates) are stored: the precision of its input feature and threshold, and the margin $m$ by which to alter the threshold value, in order to substitute it with a hardware-friendlier one.
Our flexible precision conversion module produces the new threshold values in fixed-point and integer form, for accuracy and area estimation, respectively, as mentioned above.
All (initially random) chromosomes (i.e., parent population) are subjected to the standard iterative procedure of the NSGA-II, i.e., tournament selection, simulated binary crossover and polynomial mutation.
From a combined pool of parent and children chromosomes, the non-dominated solutions are selected via fast non-dominated sorting and truncation based on crowding distance.
After the selected number of generations, our genetic optimization outputs a set of pareto-optimal approximate solutions of high hardware-efficiency and negligible classification errors.

%% file: evaluation.tex
\section{Experimental Evaluation}\label{sec:evaluation} % ~ 1 + 1/2 pages
\input{tables/baseline_table}
\begin{figure*}[t]
    \centering
    \resizebox{\textwidth}{!}{
        \includegraphics{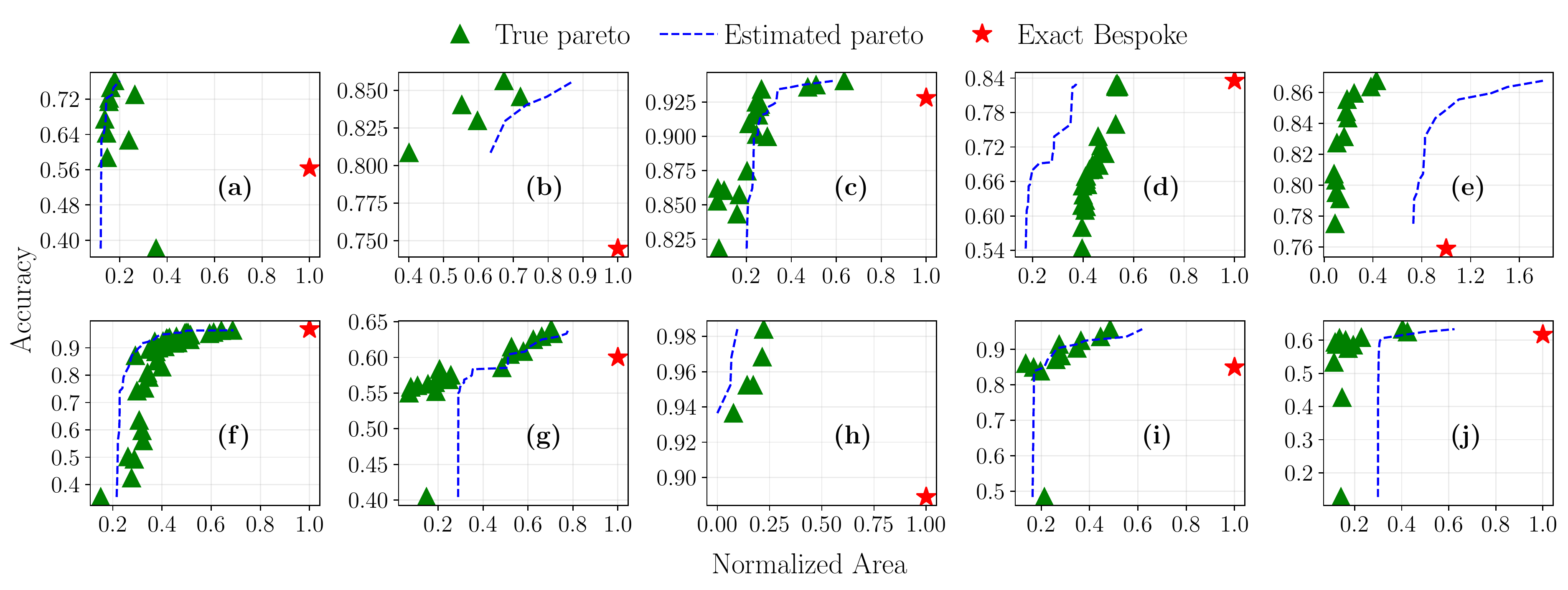}
    }
    \vspace{-10pt}
    \caption{Pareto fronts across all examined ML datasets: (a) Arrhythmia, (b) Balance, (c) Cardiotography, (d) HAR, (e) Mammographic, (f) PenDigits, (g) RedWine, (h) Seeds, (i) Vertebral, (j) WhiteWine.}
    \vspace{-5pt}
    \label{fig:pareto}
\end{figure*}
In this section, we evaluate the efficiency of our framework in generating approximate bespoke decision trees with high area and power efficiency and negligible classification accuracy loss.
We evaluate our framework over 10 datasets of the UCI ML repository~\cite{Dua:2019:datasets}, specifically: Arrhythmia, Balance, Cardiotography (Cardio), HAR, Mammographic, PenDigits, RedWine/WhiteWine, Seeds and Vertebral.
In our work, we use a fully-parallel, 8-bit area-optimized bespoke Decision Tree, like the one described in~\cite{Mubarik:MICRO:2020:printedclf} as the baseline for our experiment.
Without loss of generality, the tree depth is set so that nodes are expanded until all leaves are pure (i.e., maximum number of leafs), in order to encourage the learning of multiple decision rules but with the presence of more comparators in the tree structure.
Each exact (i.e., non-approximate) decision tree was trained using scikit-learn, with normalized training data in the interval $[0,1]$ and a random train/test split of $30\%$.
In order to avoid restricting our genetic algorithm to a confined design space by applying uniform precision, we allowed per-comparator precision to variate between $2$ and $8$ bits, whereas the margin $m$ for the threshold substitution (see Section~\ref{sec:approximations}) was set at $\pm 5$ of the original coefficient.
Thus, a variety of different mixed-precision approximate solutions per dataset can be explored and a diverse pareto front can be derived.
The resulting RTL description of the pareto-optimal bespoke Decision Trees is automatically created, by parsing the tree structure, and synthesized using Synopsys Design Compiler, at a relaxed clock (i.e., $50ms$ across all circuits).
All designs were mapped to the inkjet-printed Electrolyte Gated Transistor (EGT) library~\cite{Bleier:ISCA:2020:printedmp}, described in Section~\ref{sec:pe}.
Area measurements are directly obtained from the Design Compiler, where as power estimations are extracted using Synopsys Primetime. 
The same EDA tool flow was used for gathering all the necessary area measurements from bespoke comparators to create the Area LUT (see Section~\ref{sec:approximations}).

Table~\ref{tab:baseline} presents the characteristics of each exact bespoke Decision Tree used in our evaluation.
Note that most implemented designs feature area higher than $100mm^2$.
Additionaly, they exhibit such significant power overhead that most designs cannot be powered by existing Blue Spark printed batteries (i.e., power $<3mW$) and none can be self-powered with energy harvesters ($<0.1mW$).
Delay measurements are included for completion but are out of the scope of this work.

The time complexity of our heuristic optimization (i.e., genetic algorithm) is mainly influenced by the number of available threads and the user-defined number of generations and population size.
Thus, the execution time of a single fitness evaluation (i.e., extracting accuracy and area measurements from a candidate chromosome) establishes the bottleneck for the time complexity of the optimization procedure.
Our framework is executed purely on high level (i.e., Python) and can therefore fully exploit the inherently parallel nature of genetic algorithms.
Overall, the time complexity of the genetic algorithm is very low across all examined datasets.
Indicatively, the slowest single-chromosome evaluation had a duration of $3.08 ms$, for the HAR dataset, which is evidence of our framework's time efficiency.
Note that the execution time is proportional to the dataset difficulty.

Fig.~\ref{fig:pareto} presents the obtained pareto front for each of the examined ML classification datasets.
All presented pareto points are evaluated using the tool flow described above.
Area measurements are normalized w.r.t. the baseline values reported in Table~\ref{tab:baseline} of the exact bespoke design.
The baseline designs are also included in Fig.~\ref{fig:pareto} and are depicted by red stars.
Fig.~\ref{fig:pareto} also presents the estimated pareto front (in blue dotted lines) for each examined dataset, for the purposes of gauging the effectiveness of our heuristic optimization in discovering solutions of high area efficiency without sacrificing significant accuracy loss during the iterative process of the NSGA-II.
It can be observed that, despite a few cases (e.g., HAR, Mammographic and WhiteWine datasets), the estimated pareto front aligns very well with the actual measurements and provides a great proxy for exploring the design space of approximate bespoke Decision Trees.
Overall, the generated approximate classifiers effectively reduce the area complexity of their exact counterparts.
Each derived solution features lower area than the exact bespoke implementation and the majority of pareto-optimal solutions belong to a higher, non-dominated front than the exact classifier (i.e., at least one of the objective metrics can be improved without deteriorating the other).
Specifically, some solutions even provide higher accuracy than the exact bespoke design.
This can be explained by the fine granularity at which we scale the precision of each individual comparator in the tree structure, which acts as a regularization measure and helps produce a more efficient, less redundant classifier architecture.
On average, within a $2\%$ threshold for accuracy loss, the average area reduction w.r.t. exact classifiers achieved from by our approximate designs ranges from $5.7\times$ (for the Arrhythmia dataset) to $1.7\times$ (for the Balance dataset).
For the more demanding $1\%$ accuracy loss threshold the range is $5.7\times-1.5\times$, which still corresponds to highly efficient bespoke architectures.

\input{tables/power_table}
Our approximation framework also improves on the power-efficiency of bespoke decision trees.
Table~\ref{tab:area_power} presents an evaluation of approximate desision trees which satisfy a conservative accuracy threshold of $1\%$ (i.e., all designs in Table~\ref{tab:area_power} have higher accuracy than $0.01$ of the corresponding baseline accuracy of Table~\ref{tab:baseline}).
We included normalized area and power measurements w.r.t. the corresponding exact bespoke circuit for a better overview of the hardware efficiency of our approximate designs.
Despite of the area-driven nature of our automated framework, our approximate circuits improve upon the power consumption of exact bespoke decision trees as well.
Depending on the power-efficiency of each approximate classifier, we highlighted the measurements accordingly.
All but four generated approximate classifiers can be powered by Blue Spark printed batteries (i.e., feature power $<3mW$) and are highlighted in green.
Our most power-efficient design, for the Seeds dataset, is able to be energy-independent and powered by harvesters (i.e., features power $<0.1mW$) and is highlighted in orange.
Overall, our power-efficient approximate designs of Table~\ref{tab:area_power} provide a $3.4\times$ power reduction across all examined circuit.
Note, the respective area reduction is on average $3.2\times$, which demonstrates the multi-level benefits of our approximate design on bespoke architectures.
As can be seen, six examined approximate classifiers feature area lower than $100mm^2$.
In fact, three separate approximate bespoke classifiers (Mammographic, Seeds and Vertebral) have reported area of less than $10mm^2$, thus providing excellent candidates for printed applications.
Overall, even for very conservative classification accuracy requirements, our automated framework manages to find both area- and power-efficient solutions, which can be implemented to support a plethora of, previously infeasible, printed applications.

%% file: tables/baseline_table.tex
\begin{table}[t]
\caption{Evaluation of exact bespoke Decision Tree circuits for each examined dataset.}
\centering
{\footnotesize
\setlength{\tabcolsep}{4pt}
\begin{threeparttable}
\begin{tabular}{c|c|c|c|c|c}
	\hline
	\textbf{Dataset} & \textbf{Accuracy} & \textbf{\#Comp.\tnote{1}} & \makecell{\textbf{Delay} \\ ($ms$)} & \makecell{\textbf{Area} \\ ($mm^2$)} & \makecell{\textbf{Power} \\ ($mW$)} \\
	\hline
	\textbf{Arrhythmia} & 0.564 & 54 & 27.0 & 162.50 & 7.55 \\
	\hline
	\textbf{Balance} & 0.745 & 102 & 28.0 & 68.04 & 3.11 \\
	\hline
	\textbf{Cardio} & 0.928 & 79 & 30.4 & 178.63 & 8.12 \\
	\hline
	\textbf{HAR} & 0.835 & 178 & 33.7 & 551.08 & 26.10 \\
	\hline
	\textbf{Mammogr.} & 0.759 & 150 & 34.2 & 98.75 & 4.47 \\
	\hline
	\textbf{PenDigits} & 0.968 & 243 & 36.9 & 574.46 & 25.00 \\
	\hline
	\textbf{Redwine} & 0.600 & 259 & 38.7 & 513.84 & 22.30 \\
	\hline
	\textbf{Seeds} & 0.889 & 10 & 20.3 & 30.13 & 1.43 \\
	\hline
	\textbf{Vertebral} & 0.850 & 27 & 20.9 & 57.70 & 2.68 \\
	\hline
	\textbf{WhiteWine} & 0.617 & 280 & 49.9 & 543.12 & 23.20 \\
	\hline

\end{tabular}
\begin{tablenotes}
\item[1] Number of comparators in the design
\end{tablenotes}
\end{threeparttable}
}
\vspace{-10pt}
\label{tab:baseline}
\end{table}

%% file: tables/power_table.tex
\begin{table}[t]
\caption{Area and power evaluation for an accuracy threshold of $1\%$. Circuits highlighted in green can be powered by Blue Spark printed batteries ($<3mW$) and the design highlighted in orange can be self-powered, i.e. by an energy harvester ($<0.1mW$)}\centering
{\footnotesize
\begin{tabular}{c|c|c|c|c|c}
	\hline
	\textbf{Dataset} & \textbf{Accuracy} & \makecell{\textbf{Area} \\ ($mm^2$)} & \makecell{\textbf{Norm.} \\ \textbf{Area}} &  \makecell{\textbf{Power} \\ ($mW$)} & \makecell{\textbf{Norm.} \\ \textbf{Power}} \\
	\hline
	\rowcolor{DarkSeaGreen1!50} \textbf{Arrhythmia} & 0.67 & 22.30 & 0.137 & 1.04 & 0.138 \\
	\hline
	\rowcolor{DarkSeaGreen1!50} \textbf{Balance} & 0.81 & 27.28 & 0.401 & 1.16 & 0.372 \\
	\hline
	\rowcolor{DarkSeaGreen1!50} \textbf{Cardio} & 0.92 & 43.54 & 0.244 & 2.05 & 0.253 \\
	\hline
	\textbf{HAR} & 0.83 & 294.54 & 0.534 & 13.70 & 0.525 \\
	\hline
	\rowcolor{DarkSeaGreen1!50} \textbf{Mammogr.} & 0.81 & 8.06 & 0.082 & 0.38 & 0.084 \\
	\hline
	\textbf{PenDigits} & 0.96 & 368.48 & 0.641 & 16.10 & 0.644 \\
	\hline
	\textbf{Redwine} & 0.60 & 267.21 & 0.520 & 11.70 & 0.525 \\
	\hline
	\rowcolor{Burlywood1!50} \textbf{Seeds} & 0.94 & 2.32 & 0.077 & 0.09 & 0.064 \\
	\hline
	\rowcolor{DarkSeaGreen1!50} \textbf{Vertebral} & 0.86 & 7.84 & 0.136 & 0.38 & 0.142 \\
	\hline
	\textbf{WhiteWine} & 0.61 & 124.11 & 0.229 & 5.35 & 0.230 \\
	\hline

\end{tabular}
}
\vspace{-5pt}
\label{tab:area_power}
\end{table}

%% file: conclusion.tex
\section{Conclusion}\label{sec:conclusion}
Printed electronics have emerged as a promising alternative to silicon technologies, for application domains with conformity, low production time and low cost requirements.
However, the large feature sizes of PE elements deem the implementation of most printed ML classifiers infeasible.
In this work, we leverage approximate computing principles on bespoke architectures to push the boundaries of area and power efficiency of printed Decision Trees.
By employing a genetic algorithm to explore the design space of possible approximate bespoke architectures, we obtain pareto-optimal and area-efficient solutions, without sacrificing classification accuracy, which are suitable for ultra-resource constrained printed applications and tiny circuits.
Our evaluation procedure over 10 ML datasets demonstrates the effectivenes of approximate computing on reducing the area and power overhead of bespoke ML classifiers.